# Collective movement of charges in an electromagnetic field and collective electrodynamics forces

Yuriy A. Spirichev

The State Atomic Energy Corporation ROSATOM, "Research and Design Institute of Radio-Electronic Engineering" - branch of joint-stock company "Federal Scientific-Production Center "Production Association "Start" named after Michael V.Protsenko", Zarechny, Penza region, Russia

E-mail: yurii.spirichev@mail.ru

01.02.2020

The symmetric tensor energy-impulse of interaction of collective of electric charges with an electromagnetic field is received. A system of covariant energy and momentum conservation equations or a system of equations for the collective motion of charged particles is derived from this tensor. From this system Expressions are derived for collective electrodynamics forces. From the energy-momentum tensor derived the tensor of collective electrodynamics forces is.

**1 Introduction**
**2 EMT interaction EMF with charges and currents**
**3 Mechanical EMT**
**4 Equations of collective movement of charges**
**5 Collective electrodynamics forces**
**6 Tensor of collective electrodynamics forces**
**7 Conclusion**
**References**

**1 Introduction**

The theory of collective motion of charged particles in the electromagnetic field (EMF) is used for the theoretical explanation of many natural and laboratory electrodynamics phenomena, for solving practical problems of plasma retention, creating devices for plasma relativistic microwave electronics, plasma sources of x-ray and neutron radiation, plasma engines, and new collective methods for accelerating charged particles. The collective motion of charged particles can be divided into two boundary cases. In the first case, there is no external EMF, and their movement occurs under the action of a self-consistent EMF created by the charges themselves. In the second case, the charges are affected by a strong external EMF, the value of which significantly exceeds their self-consistent field. In this case, self-consistent EMF can be ignored and the movement of charges can only be considered under the influence of an external field. This article mainly deals with the second boundary case, since it is more General.

After replacing the external EMF with a self-consistent field of charges, and adding the field source equations, it leads to the first boundary case.

In recent years, much attention has been drawn to the problem of interaction of super-powerful ultrashort laser pulses with matter. Modern pulsed petawatt-class lasers create an energy density approaching nuclear, while the electron and ion velocities of the laser plasma are ultrarelativistic [1 - 4]. When considering these problems, it is no longer sufficient to limit ourselves to stationary, quasi-stationary, and non-relativistic approximations, since the high energy and speed of impact leads to the appearance of new synergistic effects that require their theoretical description.

Traditionally, the movement of charged particles in EMF is described by the canonical Lorentz equation, which is sometimes called the Lorentz law and put it on a par with the equations Maxwell's [5 - 7]:

$$\frac{d\mathbf{p}}{dt} = e \cdot (\mathbf{E} + \mathbf{V} \times \mathbf{B}) \quad \text{or} \quad \mathbf{F}^M = \mathbf{F}^E \qquad (1)$$

where $\mathbf{p}$ - is the momentum of a particle with a charge $\mathbf{e}$, $\mathbf{E}$ - is the electric field strength, $\mathbf{B}$ - is the magnetic field induction, and $\mathbf{V}$ - is the velocity of the particle. This equation describes the balance of mechanical $\mathbf{F}^M$ and electromagnetic $\mathbf{F}^E$ forces acting on a charged particle. To describe the movement of collective of charges in Eq. (1), a multiplier is introduced that takes into account the number of particles or the momentum $\mathbf{p}$ is replaced by the momentum density, and the charge $\mathbf{e}$ is replaced by the charge density ρ. Eq. (1) is used in most plasma models (independent particle models, magneto hydrodynamic models, kinetic models with a self-consistent field) from classical works to modern research [8 - 15].

When moving charged particles, the laws of conservation of energy and momentum must be performed. The covariant equations of these laws are four-dimensional divergences of the electromagnetic and mechanical energy-momentum tensors (EMT). Electromagnetic and mechanical systems are inextricably linked when charged particles move. The source of the energy and momentum of one of them is the drain for the other, and conversely, so you can write these four-dimensional equations in the form of equality of EMT divergences:

$$\partial^\mu T^E_{\mu\nu} = \partial^\mu T^M_{\mu\nu} \quad \text{и} \quad \partial^\nu T^E_{\mu\nu} = \partial^\nu T^M_{\mu\nu} \qquad (2)$$

where $T^E_{\mu\nu}$ and $T^M_{\mu\nu}$, respectively, the electromagnetic and mechanical EMT of the interaction of the EMF with charged particles. Thus, the EMT the physical system can be considered a generalized description of conservation laws. Then separate three-dimensional equations of conservation laws for energy and momentum, following from EMT, should be considered as a system of related equations describing the motion of charged particles. All EMT components have the dimensionality of energy density, and its derivatives have the dimensionality of force density. Therefore, the equations of conservation laws that follow from EMT can be considered as a system of equations the balance of electrodynamics and mechanical forces, similar to Eq. (1). It follows from the above that a simple equation of the balance of forces (1) is not sufficient to describe the collective movement of charges, since it does not sufficiently take into account the laws of conservation of moving collectives of charged particles.

The purpose of this work is to obtain a system of covariant equations for the collective motion of charged particles in a strong EMF, when the self-consistent EMF of charges can be neglected.

In this paper, the basis for considering collective motion and collective forces is the energy of interaction of an external EMF with electric charges in a vacuum. The generalized density of this energy is known as an invariant $I_E = \varphi \cdot \rho - \mathbf{A} \cdot \mathbf{J}$ [16, 17], where φ and $\mathbf{A}$ - are the scalar and vector EMF potentials, and ρ and $\mathbf{J}$ - are the charge density and current density vector. For this reason the EMF here will be described using a four-dimensional electromagnetic potential $\mathbf{A}_\mu(\varphi/c, i \cdot \mathbf{A})$, and charges and currents, in the form of a four-dimensional vector of current density $\mathbf{J}_\nu(c \cdot \rho, i \cdot \mathbf{J})$. Electric charges are considered point charges, and their magnetic moment and spin are not taken into account. The four-dimensional density of the mechanical pulse will describe as a four-dimensional vector $\mathbf{P}_\nu(c \cdot m, i \cdot \mathbf{p})$, where $m$ and $\mathbf{p}$ - are the mass density and momentum vector of the particles, and c - is the speed of light. The four-dimensional velocity of the particles will describe as a four-dimensional vector $\mathbf{V}_\nu(c, i \cdot \mathbf{V})$, where $\mathbf{V}$ - is the velocity vector of the particles. In the description of four-dimensional vector quantities accepted here, it is possible not to distinguish covariant and contra-variant indexes.

**2 EMT interaction EMF with charges and currents**

The EMF interaction EMT $T_{\mu\nu}^E$ with charges and currents is obtained [18] as the tensor product of the electromagnetic potential $\mathbf{A}_\mu$ by the four-dimensional current density $\mathbf{J}_\nu$:

$$T_{\mu\nu}^E = \mathbf{A}_\mu \otimes \mathbf{J}_\nu = \begin{pmatrix} \rho \cdot \varphi & \frac{1}{c} i \cdot \varphi \cdot J_x & \frac{1}{c} i \cdot \varphi \cdot J_y & \frac{1}{c} i \cdot \varphi \cdot J_z \\ i \cdot c \cdot \rho \cdot A_x & -A_x \cdot J_x & -A_x \cdot J_y & -A_x \cdot J_z \\ i \cdot c \cdot \rho \cdot A_y & -A_y \cdot J_x & -A_y \cdot J_y & -A_y \cdot J_z \\ i \cdot c \cdot \rho \cdot A_z & -A_z \cdot J_x & -A_z \cdot J_y & -A_z \cdot J_z \end{pmatrix} = \begin{bmatrix} W & \frac{1}{c} i \cdot \mathbf{S} \\ i \cdot c \cdot \mathbf{p} & t_{mn} \end{bmatrix} \quad (3)$$

where $W = \rho \cdot \varphi$ - is the density of the total electromagnetic energy per unit volume, $\mathbf{g} = \rho \cdot \mathbf{A}$ - is the density of electromagnetic momentum; $\mathbf{S} = \varphi \cdot \mathbf{J}$ - is the density of electromagnetic energy flow; $t_{mn} = -\mathbf{A}_m \otimes \mathbf{J}_n$ - three dimensional tensor of flux density of electromagnetic momentum or stress tensor. The trace of EMT (3) is a linear invariant of the electromagnetic energy density of the interaction of EMF with charges and currents, and the Lagrangian of this interaction [19]:

$$I_E = \varphi \cdot \rho - \mathbf{A} \cdot \mathbf{J} \quad (4)$$

This invariant is known in electrodynamics as the generalized energy density of the electromagnetic interaction [16] or the Schwarzschild invariant [17].

EMT (3) is unsymmetrical and can be decomposed into symmetric and antisymmetric tensors $T^E_{\mu\nu} = T^E_{(\mu\nu)}/2 + T^E_{[\mu\nu]}/2$. These tensors have the form:

$$T^E_{(\mu\nu)} = \begin{pmatrix} 2\rho \cdot \varphi & \frac{1}{c}i \cdot \varphi \cdot J_x + i \cdot c \cdot \rho \cdot A_x & \frac{1}{c}i \cdot \varphi \cdot J_y + i \cdot c \cdot \rho \cdot A_y & \frac{1}{c}i \cdot \varphi \cdot J_z + i \cdot c \cdot \rho \cdot A_z \\ \frac{1}{c}i \cdot \varphi \cdot J_x + i \cdot c \cdot \rho \cdot A_x & -2A_x \cdot J_x & -A_x \cdot J_y - A_y \cdot J_x & -A_x \cdot J_z - A_z \cdot J_x \\ \frac{1}{c}i \cdot \varphi \cdot J_y + i \cdot c \cdot \rho \cdot A_y & -A_x \cdot J_y - A_y \cdot J_x & -2A_y \cdot J_y & -A_y \cdot J_z - A_z \cdot J_y \\ \frac{1}{c}i \cdot \varphi \cdot J_z + i \cdot c \cdot \rho \cdot A_z & -A_x \cdot J_z - A_z \cdot J_x & -A_y \cdot J_z - A_z \cdot J_y & -2A_z \cdot J_z \end{pmatrix} \quad (5)$$

$$T^E_{[\mu\nu]} = \begin{pmatrix} 0 & \frac{1}{c}i \cdot \varphi \cdot J_x - i \cdot c \cdot \rho \cdot A_x & \frac{1}{c}i \cdot \varphi \cdot J_y - i \cdot c \cdot \rho \cdot A_y & \frac{1}{c}i \cdot \varphi \cdot J_z - i \cdot c \cdot \rho \cdot A_z \\ -\frac{1}{c}i \cdot \varphi \cdot J_x + i \cdot c \cdot \rho \cdot A_x & 0 & -A_x \cdot J_y + A_y \cdot J_x & -A_x \cdot J_z + A_z \cdot J_x \\ -\frac{1}{c}i \cdot \varphi \cdot J_y + i \cdot c \cdot \rho \cdot A_y & A_x \cdot J_y - A_y \cdot J_x & 0 & -A_y \cdot J_z + A_z \cdot J_y \\ -\frac{1}{c}i \cdot \varphi \cdot J_z + i \cdot c \cdot \rho \cdot A_z & A_x \cdot J_z - A_z \cdot J_x & A_y \cdot J_z - A_z \cdot J_y & 0 \end{pmatrix} \quad (6)$$

If in EMT (3) the external EMF is replaced by a self-consistent field of charges, it will describe the energy and momentum of the collective self-consistent interaction of charges with each other. From such tensor is followed by the equations of conservation of energy density and momentum, which describe the self-consistent movement of free charges in a vacuum. These equations can be used to analyze phenomena in space plasma.

### 3 Mechanical EMT

Similarly to obtaining EMT (3), we obtain the mechanical energy-momentum tensor as the tensor product of the four-dimensional charge velocity vector $\mathbf{V}_\mu(c, i \cdot \mathbf{V})$ by the four-dimensional momentum density vector $\mathbf{P}_\nu(c \cdot m, i \cdot \mathbf{p})$:

$$T^M_{\mu\nu} = \mathbf{V}_\mu \otimes \mathbf{P}_\nu = \begin{pmatrix} m \cdot c^2 & i \cdot c \cdot p_x & i \cdot c \cdot p_y & i \cdot c \cdot p_z \\ i \cdot c \cdot p_x & -V_x \cdot p_x & -V_x \cdot p_y & -V_x \cdot p_z \\ i \cdot c \cdot p_y & -V_y \cdot p_x & -V_y \cdot p_y & -V_y \cdot p_z \\ i \cdot c \cdot p_z & -V_z \cdot p_x & -V_z \cdot p_y & -V_z \cdot p_z \end{pmatrix} = \begin{bmatrix} W & \frac{1}{c}i \cdot \mathbf{S} \\ i \cdot c \cdot \mathbf{p} & t^M_{mn} \end{bmatrix} \quad (7)$$

where $W = m \cdot c^2$ - the total energy density of the unit volume of the medium; $\mathbf{S} = \mathbf{p} \cdot c^2$ - the energy flow density vector; $t^M_{mn}$ - the three-dimensional tensor of the mechanical pulse density flow or the stress tensor. The trace of EMT (7) is a four-dimensional invariant:

$$I_M = m \cdot c^2 - \mathbf{p} \cdot \mathbf{V}$$

Given that the relativistic volume has the form $Q = q_0 \cdot \sqrt{1 - \upsilon^2/c^2}$, this invariant, after the transition from the mass density m to the mass m, corresponds to the canonical expression of the relativistic Lagrange function L [15] for a free particle, known in mechanics, taken with the inverse sign:

$$\mathsf{L} = \frac{\mathsf{m} \cdot c^2}{\sqrt{1 - \upsilon^2/c^2}} - \frac{\mathsf{m} \cdot \mathbf{V}}{\sqrt{1 - \upsilon^2/c^2}} \cdot \mathbf{V} = \mathsf{W} - \mathbf{p} \cdot \mathbf{V}$$

where W - is the total energy of the particle.

From EMT (7) follows a system of equations conservation of mass density and momentum for a neutral isotropic continuous medium, in the absence of external forces:

$$\partial_t m + \nabla \cdot \mathbf{p} = 0 \qquad \text{and} \qquad \partial_t \mathbf{p} + \partial_m (\mathbf{V}_m \otimes \mathbf{p}_n) = 0 \qquad (8)$$

By converting the first equation to the form:

$$-m \cdot \partial_t \mathbf{V} = \mathbf{V}^2 \nabla m + \mathbf{V} \cdot \partial_t m \qquad (9)$$

and substituting it into the second equation, we get the equation of the balance of mechanical forces in a neutral isotropic continuous medium, in the absence of external forces, in the form:

$$-m \cdot \partial_t \mathbf{V} + \frac{1}{2} m \cdot \mathbf{V} \times \nabla \times \mathbf{V} = 0 \qquad \text{ore} \qquad -m \cdot \partial_t \mathbf{V} + \mathbf{p} \times \mathbf{\Omega} = 0 \qquad (10)$$

Here $\mathbf{\Omega}$ - is the vector of the angular velocity of rotation of the medium. The first term of the equation represents the density force of inertial Newton, and the second term represents the centrifugal force density inertia of rotation of the medium. If the medium is not neutral, then the right side of this equation, as third-party forces, must contain electromagnetic forces, similar to Eq. (1). This case will be described in the next section.

### 4 Equations of collective movement of charges

From EMT (3) and (7), in the form of their four-dimensional divergences (2), the equations of the laws of conservation of electromagnetic energy and momentum follow. Taking into account symmetry of the tensor $T_{\mu\nu}^M$ it is possible to write down the general equation in the form (here and further it is possible not to distinguish covariant and contra-variant indexes):

$$\partial_\mu T_{\mu\nu}^E + \partial_\nu T_{\mu\nu}^E = 2 \partial_\mu T_{\mu\nu}^M$$

Taking into account decomposition $T_{\mu\nu}^E = T_{(\mu\nu)}^E / 2 + T_{[\mu\nu]}^E / 2$, we will record this equation as:

$$(\partial_\mu T_{(\mu\nu)}^E + \partial_\mu T_{[\mu\nu]}^E + \partial_\nu T_{(\mu\nu)}^E + \partial_\nu T_{[\mu\nu]}^E)/2 = 2 \partial_\mu T_{\mu\nu}^M$$

Given, that $\partial_\mu T_{(\mu\nu)}^E = \partial_\nu T_{(\mu\nu)}^E$, a $\partial_\mu T_{[\mu\nu]}^E = -\partial_\nu T_{[\mu\nu]}^E$, we will get conservation equations in the form:

$$\partial_\mu T_{(\mu\nu)}^E / 2 = \partial_\mu T_{\mu\nu}^M \qquad (11)$$

Eq. (11) shows that the antisymmetric part (6) of the unsymmetrical EMT (3) is not involved in the energy exchange with the mechanical system and only the symmetric part (5) should be considered further. This is to be expected, since the mechanical EMT (7) is symmetrical. Let's write Eq. (11) in expanded form:

$$\frac{1}{c}\partial_t(\varphi\cdot\rho)+(\nabla\cdot(c\cdot\rho\cdot\mathbf{A})+\nabla\cdot(\frac{1}{c}\varphi\cdot\mathbf{J}))/2=c\cdot(\partial_t m+\nabla\cdot\mathbf{p}) \quad (12)$$

$$(\frac{1}{c^2}\partial_t(\varphi\cdot\mathbf{J})+\partial_t(\rho\cdot\mathbf{A})+\partial_m(\mathbf{A}_m\otimes\mathbf{J}_n)+\partial_m(\mathbf{A}_n\otimes\mathbf{J}_m))/2=\partial_t\mathbf{p}+\partial_m(\mathbf{V}_m\otimes\mathbf{p}_n) \quad (13)$$

Eqs. (12) and (13) are a complete relativistically covariant system of equations describing the collective motion of charged particles in EMF. This system replaces the balance of forces Eq. (1). In the system (12)-(13), the external EMF $\mathbf{A}_\mu(\varphi/c, i\cdot\mathbf{A})$ is considered known, and the values are determined $\mathbf{J}_\nu(c\cdot\rho, i\cdot\mathbf{J})$, $\mathbf{P}_\nu(c\cdot m, i\cdot\mathbf{p})$, $\mathbf{V}_\nu(c, i\cdot\mathbf{V})$. Since the number of unknowns in the system (12)-(13) exceeds the number of equations, it is undefined. To reduce the number of unknown and determine the system you need to consider that $\mathbf{J}=\rho\cdot\mathbf{V}$, $\mathbf{P}=m\cdot\mathbf{V}$, $\rho/m=const$.

After performing differentiation in Eqs. (12)-(13) and using known vector identities, we write these equations in expanded form:

$$[\frac{1}{c}\varphi\cdot\partial_t\rho+\frac{1}{c}\rho\cdot\partial_t\varphi+(c\cdot\rho\cdot\nabla\cdot\mathbf{A}+c\cdot\mathbf{A}\cdot\nabla\rho+\frac{1}{c}\varphi\cdot\nabla\cdot\mathbf{J}+\frac{1}{c}\mathbf{J}\cdot\nabla\varphi)/2]\frac{\mathbf{V}}{c}= \\ =m\cdot\partial_t\mathbf{V}+(\mathbf{V}\cdot\partial_t m+\mathbf{V}^2\cdot\nabla m) \quad (14)$$

$$(\frac{1}{c^2}\varphi\cdot\partial_t\mathbf{J}+\frac{1}{c^2}\mathbf{J}\cdot\partial_t\varphi+\rho\cdot\partial_t\mathbf{A}+\mathbf{A}\cdot\partial_t\rho+\mathbf{J}\cdot\nabla\cdot\mathbf{A}+\mathbf{A}\cdot\nabla\cdot\mathbf{J}+\nabla(\mathbf{A}\cdot\mathbf{J})- \\ -\mathbf{A}\times\nabla\times\mathbf{J}-\mathbf{J}\times\nabla\times\mathbf{A})/2=3m\cdot\partial_t\mathbf{V}+(\mathbf{V}\partial_t m+\mathbf{V}^2\nabla m)-2m\mathbf{V}\times\Omega \quad (15)$$

In the right part of Eq. (15), supersede the brackets with their expression taken from Eq. (14) and transfer the electrodynamics forces to the right part of the equation, and the mechanical forces to the left part of the equation:

$$-m\cdot\partial_t\mathbf{V}+m\cdot\mathbf{V}\times\Omega=[\frac{1}{c}\varphi\cdot\partial_t\rho+\frac{1}{c}\rho\cdot\partial_t\varphi+(c\cdot\rho\cdot\nabla\cdot\mathbf{A}+c\cdot\mathbf{A}\cdot\nabla\rho+\frac{1}{c}\varphi\cdot\nabla\cdot\mathbf{J}+\frac{1}{c}\mathbf{J}\cdot\nabla\varphi)/2]\frac{\mathbf{V}}{2c}+ \\ +(-\frac{1}{c^2}\varphi\cdot\partial_t\mathbf{J}-\frac{1}{c^2}\mathbf{J}\cdot\partial_t\varphi-\rho\cdot\partial_t\mathbf{A}-\mathbf{A}\cdot\partial_t\rho-\mathbf{J}\cdot\nabla\cdot\mathbf{A}-\mathbf{A}\cdot\nabla\cdot\mathbf{J}-\nabla(\mathbf{A}\cdot\mathbf{J})+\mathbf{A}\times\nabla\times\mathbf{J}+\mathbf{J}\times\nabla\times\mathbf{A})/4 \quad (16)$$

Using a known vector identity $\nabla(\mathbf{A}\cdot\mathbf{J})=\nabla\times(\mathbf{A}\times\mathbf{J})+\mathbf{A}\times\nabla\times\mathbf{J}+\mathbf{J}\times\nabla\times\mathbf{A}+(\mathbf{A}\cdot\nabla)\mathbf{J}+(\mathbf{J}\cdot\nabla)\mathbf{A}$, Eq. (16) is written as:

$$-m\cdot\partial_t\mathbf{V}+m\cdot\mathbf{V}\times\Omega=[\frac{1}{c}\varphi\cdot\partial_t\rho+\frac{1}{c}\rho\cdot\partial_t\varphi+(c\cdot\rho\cdot\nabla\cdot\mathbf{A}+c\cdot\mathbf{A}\cdot\nabla\rho+\frac{1}{c}\varphi\cdot\nabla\cdot\mathbf{J}+\frac{1}{c}\mathbf{J}\cdot\nabla\varphi)/2]\frac{\mathbf{V}}{2c}+ \\ +(-\frac{1}{c^2}\varphi\cdot\partial_t\mathbf{J}-\frac{1}{c^2}\mathbf{J}\cdot\partial_t\varphi-\rho\cdot\partial_t\mathbf{A}-\mathbf{A}\cdot\partial_t\rho-\mathbf{J}\cdot\nabla\cdot\mathbf{A}-\mathbf{A}\cdot\nabla\cdot\mathbf{J}-(\mathbf{A}\cdot\nabla)\mathbf{J}-(\mathbf{J}\cdot\nabla)\mathbf{A})/4 \quad (17)$$

This equation is a complete equation of the balance of mechanical and electrodynamics forces. Its left part corresponds to the left part of Eq. (10) of the balance of mechanical forces in an isotropic neutral continuous medium and contains the Newton's inertia force and the centrifugal inertia force of the medium's rotation.

If in the system of Eqs. (12)-(13) and in Eq. (17), the external EMF is replaced by a self-consistent field of charges, then they will describe the density of energy, momentum, and forces of the collective interaction of charges with each other, as well as their self-consistent motion. For this case, the system of Eqs. (12)-(13) must be supplemented with equations describing the relationship of electric charges and currents with their own EMF. It is obvious that in the case of self-consistent EMF, the equations of motion become nonlinear.

**5 Collective electrodynamics forces**

Usually, when describing electromagnetic forces in a vacuum, EMF tensions are used in the form of $\mathbf{E}$ and $\mathbf{B}$. The use of the electromagnetic potential $\mathbf{A}_\mu(\varphi/c, i \cdot \mathbf{A})$ that characterizes energy makes it possible to present these forces in a systematic and more detailed way, which facilitates their analysis.

Consider the electrodynamics forces included in the right side of the balance of forces Eq. (17). All electrodynamics forces can be divided into two groups. The first group includes forces that depend on the rate of change in EMF in time and space, and the second group includes forces that depend on the rate of change in the density of charges and currents in time and space. The forces belonging to the first group are shown in table 1, and the forces belonging to the second group are shown in table 2.

Table 1. Forces that depend on the rate of change in EMF

| 1) $\frac{1}{c}\rho \cdot \partial_t \varphi$ | 2) $-\frac{1}{c^2}\mathbf{J} \cdot \partial_t \varphi$ | 3) $-\rho \cdot \partial_t \mathbf{A}$ | 4) $\frac{1}{c}\mathbf{J} \cdot \nabla \varphi$ | 5) $c \cdot \rho \cdot \nabla \cdot \mathbf{A}$ | 6) $-\mathbf{J} \cdot \nabla \cdot \mathbf{A}$ | 7) $-(\mathbf{J} \cdot \nabla)\mathbf{A}$ |

Table 2. Forces that depend on the rate of change in charge density and currents

| 1) $\frac{1}{c}\varphi \cdot \partial_t \rho$ | 2) $-\frac{1}{c^2}\varphi \cdot \partial_t \mathbf{J}$ | 3) $-\mathbf{A} \cdot \partial_t \rho$ | 4) $\frac{1}{c}\varphi \cdot \nabla \cdot \mathbf{J}$ | 5) $c \cdot \mathbf{A} \cdot \nabla \rho$ | 6) $-\mathbf{A} \cdot \nabla \cdot \mathbf{J}$ | 7) $-(\mathbf{A} \cdot \nabla)\mathbf{J}$ |

A comparison of the forces shown in tables 1 and 2 shows their complete symmetry with respect to EMF and the density of charges and currents. This proves that the description of collective electrodynamics forces using the electromagnetic potential $\mathbf{A}_\mu$ is more preferable than their description using EMF tensions, which do not allow describing these forces in full. Since tables 1 and 2 are completely symmetrical, it is sufficient to consider only the electrodynamics forces shown in table 1.. These forces can be divided into dynamic forces 1, 2, 3, the magnitude of which depends on the rate of change of EMF in time and stationary forces 4-7, the magnitude of which depends on the rate of change of EMF in three-dimensional space. The forces 1, 4, 5 are scalar, the other forces are vector. Scalar forces are independent of direction and are similar to gas-hydraulic pressure/compression forces in a continuous medium. Among the vector forces of table 1, the force 3 is well known, which can be written as $f_3 = -\rho \cdot \partial_t \mathbf{A} = \rho \cdot \mathbf{E}_V$, where $\mathbf{E}_V$ - vortex electric field. Table 1 shows that the equations of conservation of energy density and momentum (12)-(13), as well as the equation of the balance of forces (17) does not include the potential Coulomb force $f_K = -\rho \cdot \nabla \varphi = \rho \cdot \mathbf{E}_P$. This indicates that the potential electric field does not play a role in solving dynamic plasma problems, and only the vortex electric field should be taken into account $\mathbf{E}_V = -\partial_t \mathbf{A}$. A comparison of Eqs. (16) and (17) shows that the Ampere force $f_A = \mathbf{J} \times \nabla \times \mathbf{A} = \mathbf{J} \times \mathbf{B}$, where $\mathbf{B}$ – magnetic induction is part of the more General collective electrodynamic force 7. This force can be called the generalized Ampere force $f_{GA} = -(\mathbf{J} \cdot \nabla)\mathbf{A}$.

From the system of Eqs. (12)-(13), it follows that the collective movement of free electric charges is complex, depending not only on the four-dimensional dynamics of EMF, but also on the rate of change in the density of charges and currents. This four-dimensional dynamic results in the forces shown in table 2. In well-known theoretical works on plasma physics, these collective electrodynamics forces associated with the rate of change in the density of charges and currents are not taken into account.

**6 Tensor of collective electrodynamics forces**

The components of EMT (3) have an energy dimension, and their four-dimensional derivatives have a force dimension. Then a full generalized system description of electrodynamics forces can be considered a third-rank tensor, which is a four-dimensional derivative of the EMT $F^E_{\lambda\mu\nu} = \partial_\lambda T^E_{\mu\nu}$. The components of this tensor can formally be considered a single four-dimensional tensor force, whose components are related by the corresponding compatibility and continuity equations. However, the components of this tensor can also be represented as separate scalar and vector forces. The tensor of electrodynamics forces $F^E_{\lambda\mu\nu}$ can be written as the sum of two tensor products:

$$F^E_{\lambda\mu\nu} = \partial_\lambda T^E_{\mu\nu} = \partial_\lambda (\mathbf{A}_\mu \otimes \mathbf{J}_\nu) = \partial_\lambda \mathbf{A}_\mu \otimes \mathbf{J}_\nu + \mathbf{A}_\mu \otimes \partial_\lambda \mathbf{J}_\nu \qquad (18)$$

From this expression, it can be seen that all electrodynamics forces, similar to the forces shown in tables 1 and 2, can be divided into two groups. The first group includes forces that depend on the rate of change of the electromagnetic potential $\partial_\lambda \mathbf{A}_\mu$ at a constant current density $\mathbf{J}_\nu$ (the first term (18)). The second group includes forces that depend on the rate of change in the four-dimensional current $\partial_\lambda \mathbf{J}_\nu$ density $\mathbf{A}_\mu$ at constant (the second term (18)). These tensors of the four-dimensional rate of change of the electromagnetic potential $\mathcal{A}_{\lambda\mu} = \partial_\lambda \mathbf{A}_\mu$ and current density $J_{\lambda\nu} = \partial_\lambda \mathbf{J}_\nu$ included in the tensor (18) have the form:

$$\mathcal{A}_{\lambda\mu} = \partial_\lambda A_\mu = \begin{pmatrix} \frac{1}{c^2}\partial_t\varphi & \frac{1}{c}i\cdot\partial_t A_x & \frac{1}{c}i\cdot\partial_t A_y & \frac{1}{c}i\cdot\partial_t A_z \\ -\frac{1}{c}i\cdot\partial_x\varphi & \partial_x A_x & \partial_x A_y & \partial_x A_z \\ -\frac{1}{c}i\cdot\partial_y\varphi & \partial_y A_x & \partial_y A_y & \partial_y A_z \\ -\frac{1}{c}i\cdot\partial_z\varphi & \partial_z A_x & \partial_z A_y & \partial_z A_z \end{pmatrix} \quad J_{\lambda\nu} = \partial_\lambda J_\mu = \begin{pmatrix} \partial_t\rho & \frac{1}{c}i\cdot\partial_t J_x & \frac{1}{c}i\cdot\partial_t J_y & \frac{1}{c}i\cdot\partial_t J_z \\ -ci\cdot\partial_x\rho & \partial_x J_x & \partial_x J_y & \partial_x J_z \\ -ci\cdot\partial_y\rho & \partial_y J_x & \partial_y J_y & \partial_y J_z \\ -ci\cdot\partial_z\rho & \partial_z J_x & \partial_z J_y & \partial_z J_z \end{pmatrix}$$

These tensors have an independent meaning. Of them, in the form of divergence equations follow, respectively, the four-dimensional movement equations EMF, in the absence of electric charges, and the equations of four-dimensional motion of charges, in the absence of EMF. From the linear invariant of the tensor $\partial_\nu A_\mu$ follows the Lorentz calibration condition, and from the linear invariant of the tensor $\partial_\nu J_\mu$ follows the continuity equation of the current density $\partial_t\varphi/c^2 + \nabla \cdot \mathbf{A} = 0$ (if these invariants are equal to zero). Thus, these important expressions in electrodynamics are derived from the EMT (3) interaction of electric charges with EMF and the force tensor (18). In this case, the Lorentz calibration condition can be given a physical interpretation in the form of a description of the four-dimensional expansion/compression of the EMF. A more detailed discussion of this issue is not relevant to the subject of this article. The components of the tensor (18) describe all collective electrodynamics forces, including the potential Coulomb force that is not comprised in Eq. (17).

Tensors $\mathcal{A}_{\lambda\mu}$ and $J_{\lambda\nu}$ are unsymmetrical and can be decomposed into symmetric and antisymmetric tensors. Then the tensor of electrodynamics forces (18) can be written as four tensor products:

$$F^E_{\lambda\mu\nu} = (\mathcal{A}_{(\lambda\mu)} \otimes \mathbf{J}_\nu + \mathcal{A}_{[\lambda\mu]} \otimes \mathbf{J}_\nu + \mathbf{A}_\mu \otimes J_{(\lambda\nu)} + \mathbf{A}_\mu \otimes J_{[\lambda\nu]})/2 \qquad (19)$$

Symmetric tensors can be considered as a description of four-dimensional deformations $\mathbf{A}_\mu$ and $\mathbf{J}_\nu$, and antisymmetric tensors as a description of their four-dimensional rotations. Therefore, based on the decomposition (19), the electrodynamics forces can also be divided into four-dimensional deformation forces and four-dimensional rotation forces.

### 6 Conclusion

The use of electromagnetic potential $\mathbf{A}_\mu$ instead of EMF stresses in the form of $\mathbf{E}$ and $\mathbf{B}$ allows us to obtain a systematic and symmetrical description of collective electrodynamics forces with respect to charges and currents. When considering collective electrodynamics phenomena in strong fields, the Lorentz Eq. (1) must be replaced by a system of Eqs. (12)-(13) or Eq. (17). This system of equations for the collective movement of electric charges is relativistically covariant and describes the appearance of a number of collective electrodynamics forces that are not currently taken into account in the analysis of electrodynamics phenomena.

The system of Eqs. (12)-(13), after replacing the external EMF with the self-consistent field of the charges themselves, is a covariant equation of the collective self-consistent motion of free charged particles. For this case, the system (12)-(13) must be supplemented with equations describing the relationship of electric charges and currents with their own EMF.

A complete generalized system description of collective electrodynamics forces is the third-rank tensor, which is a four-dimensional derivative of the EMT $F^E_{\lambda\mu\nu} = \partial_\lambda T^E_{\mu\nu}$. Since all the components of the force tensor are connected to each other by the compatibility and continuity equations known in the continuum theory, they can be considered as one multi-component tensor force.

The expression of the Lorentz calibration condition for EMF and the continuity equation for the current density follow from the EMT $T^E_{\mu\nu}$ and the force tensor $F^E_{\lambda\mu\nu}$.